\documentclass[unnumsec,webpdf,contemporary,large]{oup-authoring-template}%
\usepackage{xurl}
\usepackage[mathlines, switch]{lineno}
\usepackage{bm}
\begin{document}

\journaltitle{Pre-print submitted to North American Journal of Fisheries Management}
\DOI{10.48550/arXiv.2503.17293}
\copyrightyear{2025}
\appnotes{Pre-print}

\firstpage{1}

\title[Forecasting harvest of a recreational fishery]{Leveraging statistical models to improve pre-season forecasting and in-season management of a recreational fishery}

\author[1,$\ast$]{A. Challen Hyman\ORCID{0000-0001-9493-2092}}
\author[1]{Chloe Ramsay}
\author[2]{Tiffanie A. Cross}
\author[2]{Beverly Sauls}
\author[1]{Thomas K. Frazer}

\authormark{Hyman et al. 2025}

\address[1]{\orgname{College of Marine Science, University of South Florida}, \orgaddress{\street{140 7th Avenue South}, \postcode{33701}, \state{Florida}, \country{USA}}}

\address[2]{\orgname{Florida Fish and Wildlife Conservation Commission, Fish and Wildlife Research Institute}, \orgaddress{\street{100 Eighth Avenue SE}, \postcode{33701}, \state{Florida}, \country{USA}}}

\corresp[$\ast$]{Corresponding author. \href{email:email-id.com}{achyman@usf.edu}}

\abstract{Effective management of recreational fisheries requires accurate forecasting of future harvests and real-time monitoring of ongoing harvests. Traditional methods that rely on historical catch data to predict short-term harvests can be unreliable, particularly if changes in management regulations alter angler behavior. In contrast, statistical modeling approaches can provide faster, more flexible, and potentially more accurate predictions, enhancing management outcomes. In this study, we developed and tested models to improve predictions of Gulf of Mexico gag harvests for both pre-season planning and in-season monitoring. Our best-fitting model outperformed traditional methods (i.e., estimates derived from historical average harvest) for both cumulative pre-season projections and in-season monitoring. Notably, our modeling framework appeared to be more accurate in more recent, shorter seasons due to its ability to account for effort compression. A key advantage of our framework is its ability to explicitly quantify the probability of exceeding harvest quotas for any given season duration. This feature enables managers to evaluate trade-offs between season duration and conservation goals. This is especially critical for vulnerable, highly targeted stocks. Our findings also underscore the value of statistical models to complement and advance traditional fisheries management approaches.}

\maketitle

\section{Introduction}
One of the primary management strategies in recreational fisheries is the use of temporal harvest restrictions to reduce fishing pressure. Annual fishing quotas, determined in large part by stock assessments, are used to help establish the duration of the fishing season \citep{methot2014implementing}. The actual duration of the recreational fishing season is based on projections of when the quota will be exceeded. Therefore, the accuracy of these projections, along with in-season monitoring as the season progresses, is crucial to minimizing the risk of overfishing and ensuring sustainable fisheries.

Both pre- and in-season harvest projections are typically estimated using historical, temporally indexed harvest information from previous years \citep{carruthers2014evaluating}. Recreational fishing season closure dates are often established before the season begins to provide private anglers and for-hire businesses with a clear expectation of the total season duration. This approach relies on projections for when the recreational quota will be met before any harvest data from the present year are available \citep{farmer2015forecasting, farmer2020forecasting}. Additionally, temporally indexed recreational harvest estimates are seldom available on timelines required for in-season management \citep{carter2015nowcasting, MRIPData2021}. Due to the substantial number of participants in many recreational fisheries, recreational harvest is often estimated using sampling data from dockside interviews and mail surveys of effort, followed by survey-based extrapolation methods \citep{MRIPData2021, MRIP2023}. The time required to implement this methodology, especially the mail survey, results in estimate reporting lags of up to several months after sampling. Consequently, managers are often forced to rely on limited in-season and past harvest information to formulate and update harvest forecasts as the season progresses.
 
 Unfortunately, past harvest data can be an unreliable predictor of future harvest rates when regulations change \citep{carter2015nowcasting}. Angler behaviors, including decisions about whether to fish and which species to target, are heavily influenced by fishing regulations \citep{farmer2020forecasting, Hyman2024Effort}. Anglers often focus on popular species that are currently open for harvest and may concentrate their efforts on a particular species if its season is shortened \citep{abbott2018status, trudeau2022lower}. When harvests exceed projections based on past estimates, managers may need to implement post-hoc accountability measures, such as reducing the quota for the subsequent year \citep[i.e., a payback;][]{methot2014implementing}. Participants in the recreational fishery may perceive these measures as punitive, despite their adherence to regulations, potentially leading to decreased future compliance and undermining management integrity \citep{cowan2011faith, cowan2011red}. Therefore, techniques that bridge the gap between harvest data availability and real-time management needs are valuable tools for ensuring sustainable fisheries and promoting recreational angler satisfaction.

 Modeling frameworks leveraging relationships between readily available social, environmental, and management variables and harvest have demonstrable success in improving the accuracy of both pre-season projection and in-season monitoring estimates. For example, a suite of sophisticated modeling approaches has been applied to reliably forecast landings from a broad range of species, including Gulf of Mexico (Gulf) vermilion snapper (\textit{Rhomboplites aurorubens}) and gray snapper (\textit{Lutjanus griseus}) \citep[][]{farmer2015forecasting}, Gulf red snapper (\textit{Lutjanus campechanus})\citep[][]{farmer2020forecasting}, Atlantic menhaden (\textit{Brevoortia tyrannus})\citep[][]{hanson2006forecasting}, Pacific giant octopus (\textit{Enteroctopus dofleini}) \citep[][]{nagano2023predicting}, and the Indian oil sardine (\textit{Sardinella longiceps}) \citep[][]{holmes2021improving}. Such methods can be used to set season durations \citep{farmer2020forecasting} or `nowcast' in-season catch rates \citep{carter2015nowcasting} using combinations of fishing regulations, socioeconomic factors, weather conditions, and internet search traffic. In fact, these modeling approaches can represent considerable improvements to more typical projections primarily based on past harvest data \citep{carter2015nowcasting}. 
 
 The accuracy gap between harvest projections based on statistical models and those derived from traditional methods may be most pronounced during stock rebuilding periods when recreational quotas are low and seasons are significantly shortened. Managers typically reduce season durations to alleviate fishing pressure on vulnerable stocks. However, anglers may respond by intensifying their effort within the limited fishing window, leading to increased harvest rates \citep{abbott2018status, farmer2020forecasting}. This phenomenon, known as ``effort compression'', has been widely documented in the recreational Gulf red snapper fishery \citep[e.g.,][]{powers2016estimating, powers2018compression, topping2019comparison, farmer2020forecasting, Hyman2024Effort}, where federal season reductions from 194 days in 2005 to just nine days in 2014 resulted in exponentially higher catch rates for both private and for-hire sectors \citep[see Fig. 4 in][]{farmer2020forecasting}. Effort compression is most pronounced during the shortest seasons, typically when a stock is overexploited. Consequently, failing to account for this behavior can lead to substantial harvest overages precisely when the stock is most vulnerable. While traditional projections based on historical harvest averages cannot inherently account for these behavioral shifts, statistical modeling approaches can explicitly incorporate season duration effects \citep{Hyman2024Effort, farmer2020forecasting}, theoretically leading to more accurate forecasts and improved management outcomes. However, such frameworks have rarely been applied in management settings \citep{farmer2020forecasting}.

 In this study, we apply a regression framework to forecast harvest for the recreational sector component of the Gulf gag (\textit{Mycteroperca microlepis}) stock. Gulf gag are highly targeted by recreational anglers \citep{Hyman2024Effort} and the gag fishery is currently in a rebuilding plan \citep{Amendment56}. Recreational catch-per-unit-effort (CPUE) has declined in recent years \citep{GAGSEDAR72} and stock status is currently considered impaired \citep[i.e., overfishing is occurring and the stock is overfished; ][]{GAGSEDAR72, Gag_INTERIM, Amendment56}. In response, the Gulf of Mexico Fishery Management Council (GMFMC) reduced the recreational gag annual catch limit (ACL) by 80\% and reduced the season duration from 214 days in 2022 to 49 days in 2023 \citep{Amendment56} and 15 days in 2024 \citep{NOAA2024GagSeason}. Projections of season duration based on harvest rates from previous years underestimated harvest in the substantially shorter season, and the ACL was exceeded in both 2023 \citep{Amendment56, NOAA2024GagSeason} and with preliminary 2024 estimates, suggesting intensified fishing effort within the shortened time frame. Consequently, more precise projection methodologies that (1) incorporate changes in harvest rates driven by temporal restrictions (i.e., effort compression) and environmental conditions and (2) quantify the risk of exceeding the harvest quota could enhance future fisheries management decisions. We developed a set of statistical models to generate monthly estimates of harvest rates (pounds per day open) using management, socioeconomic, environmental, and seasonal predictors. We also demonstrate this framework may be used to generate pre-season forecasts and in-season updates of when a hypothetical quota will be met. 

\section{Methods}
\subsection{Data sources}\label{survey}
We compiled \href{https://www.st.nmfs.noaa.gov/st1/recreational/MRIP_Survey_Data/}{recreational gag harvest data} for the for-hire sector from the National Marine Fisheries Service (NMFS) Marine Recreational Information Program (MRIP) \href{https://www.fisheries.noaa.gov/recreational-fishing-data/hire-survey-glance}{For-Hire Survey (FHS)} combined with data from the \href{https://www.fisheries.noaa.gov/recreational-fishing-data/access-point-angler-intercept-survey-glance}{Angler Point Access Intercept Survey (APAIS)} and private recreational data from the \href{https://myfwc.com/fishing/saltwater/recreational/state-reef-fish-survey/}{Florida Fish and Wildlife (FWC) State Reef Fish Survey (SRFS)}. This approach aligns with the data sources currently used for gag management, as both the GMFMC and NOAA NMFS monitor recreational gag landings using SRFS and FHS/APAIS data. 

Since 2023, the annual recreational gag quota has been determined based on the most recent stock assessment \citep{GAGSEDAR72} in conjunction with a stock rebuilding framework \citep{Amendment56} both of which fundamentally rely on recreational landings from the private fleet (represented by SRFS) and the for-hire fleet (represented by FHTS). Aggregated monthly landings from these surveys are used to monitor quota attainment. We applied similar aggregation methods to construct our predictive models, with two exceptions. First, although MRIP collects data from shore-based anglers and includes shore harvest in the recreational quota, we excluded shore mode from our analysis due to its historically negligible contribution to gag landings and the high uncertainty in its estimates \citep{MRIP2024_gulf}. Second, our analysis focused on the west coast of Florida only (excluding the Florida Keys) rather than the Gulf as a whole, as more than 98\% of all Gulf gag landings have occurred in Florida since 2011 \citep{MRIP2024_gulf,Amendment56}.

The State Reef Fish Survey (SRFS) was established in 2015 (originally named the Gulf Reef Fish Survey [GRFS]) to improve harvest estimates for offshore species that are less frequently targeted and caught. The survey focuses on reef-associated species, including gag, red grouper (\textit{Epinephelus morio}), gray triggerfish (\textit{Balistes capriscus}), red snapper, vermilion snapper, greater amberjack (\textit{Seriola dumerili}), hogfish (\textit{Lachnolaimus maximus}), mutton snapper (\textit{Lutjanus analis}), and yellowtail snapper (\textit{Ocyurus chrysurus}). The SRFS estimates CPUE using a combination of data collected on these reef-associated species from the APAIS and data collected from SRFS intercepts, which are conducted at public boat ramps where offshore fishermen are more commonly encountered. The SRFS mail survey is sent to a stratified, random sample of anglers with the free state reef fish angler designation on their Florida saltwater license, asking them to report their fishing activity over the past month. The combination of increased dockside sampling at offshore-access points, a larger sample size, and a more targeted mail survey frame allows the SRFS to produce more precise harvest estimates for reef fish compared to the MRIP survey, which has a broader scope of all saltwater fish species.

The NOAA MRIP For-Hire Survey (FHS) estimates recreational catch and effort for charter fishing trips in U.S. waters. This survey targets federally permitted for-hire vessels, including charter boats and headboats, which operate with licensed captains and provide fishing opportunities for recreational anglers. Catch and effort estimates are derived from a combination of APAIS dockside interviews, electronic logbooks, and randomly assigned telephone surveys, which are used to calculate total catch for the sector. 

Both SRFS and FHS use dockside sampling to collect catch-per-trip data. In these interviews, trained samplers record details from anglers about their completed fishing trips, including the number and species of fish caught, the size of landed fish, and fishing effort (e.g., hours fished). These catch data are then combined with effort estimates from the mail or telephone surveys to calculate total harvest. Statistical methods are applied to account for sampling variability and potential reporting biases, ensuring robust estimates for fisheries management. To standardize harvest rates from SRFS and FHTS in Florida, we aggregated monthly gag harvest estimates across months with varying open seasons. We then divided each monthly estimate by the number of days open to fishing to calculate daily harvest rates. To avoid division by zero in months when harvest was closed, we added one additional day to the denominator.

\subsection{Predictors considered}\label{Predictors}
We considered multiple regulatory, socioeconomic, and environmental predictors to model patterns in recreational gag harvest rates (Table \ref{tab1}). Fundamentally, we expected gag harvest rates to be positively related to whether a month was open to gag harvest, as anglers would presumably target gag when the season is open \citep{Hyman2025Catch}. However, gag temporal regulations along the Gulf Coast of Florida are complex and have historically differed spatially, requiring careful consideration of how to translate these regulations into relevant predictors. From 2012 to 2022, a four-county special season for gag in Florida applied to Franklin, Wakulla, Jefferson, and Taylor counties. During this period, state waters in these counties were opened for recreational gag harvest from April 1 through June 1 or June 30, depending on the year \citep{Hyman2024Effort}. This special season allowed for local adjustments to manage the species effectively and support angling opportunities. For all Gulf state and federal waters, gag harvest was generally open from June 1 to December 31. To account for spatial differences in gag temporal restrictions, we created three dummy variables denoting whether the four-county season (denoted ``$\text{Special}_{Gag}$'') or the remainder of Gulf state and federal waters (denoted ``$\text{Full}_{Gag}$'') was open, or neither (denoted ``$\text{Closed}_{Gag}$''). Furthermore, for the hurdle component of our regression model (see Analysis section below), we introduced a fourth binary variable, ``$\text{Open}_{Gag}$'', which indicated whether either season was open. This was based on the expectation that the probability of observing zero gag harvested per day would depend on whether gag harvest anywhere was permitted during that period.

\begin{table*}[ht]
\caption{Symbology and descriptions of each predictor considered in SRFS gag harvest models. \label{tab1}}
\tabcolsep=0pt%%
\begin{tabular*}{\textwidth}{@{\extracolsep{\fill}}p{2cm} p{6cm} p{6cm}@{\extracolsep{\fill}}}
\toprule%
\textbf{Variable} &\textbf{Description} &\textbf{Data source and references}\\ \hline
 \\
  $\text{Closed}_{Gag}$& Gag harvest rate when the season is closed to fishing, taken to be the reference intercept & Florida Fish and Wildlife Conservation Commission\\ 
  $\text{Special}_{Gag}$& Binary variable denoting whether the four-county ``special'' gag season is open & Florida Fish and Wildlife conservation Commission\\ 
  $\text{Full}_{Gag}$& Binary variable denoting whether the gag season is open statewide & Florida Fish and Wildlife  Conservation Commission\\
  $\text{Open}_{Gag}$& Binary variable denoting whether the gag season is open at all & Florida Fish and Wildlife Conservation Commission\\
  $\text{Season}_{Gag}$& $\ln$ Recreational gag season duration in Florida & Florida Fish and Wildlife Conservation Commission\\ 
  $\text{Past}$& $\ln$ index of abundance taken as the average gag harvest in a given month based on the most recent three years in which that month has been open & Marine Recreational Information Program and Florida State Reef Fish survey\\ 
  $\text{Days}_{RS}$& Fraction of days in a month open to red snapper & Florida Fish and Wildlife Conservation Commission\\ 
  $\sin_{12}$& Annual sine term & $\sin_{12,t} = \sin(\frac{2\pi \cdot t}{12})$ where $t$ is a given month\\ 
  $\cos_{12}$& Annual cosine term & $\cos_{12,t} = \cos(\frac{2\pi \cdot t}{12})$ \\ 
  $\sin_{6}$& Semi-annual sine term & $\sin_{6,t} = sin(\frac{2\pi \cdot t}{6})$ \\ 
  $\cos_{6}$& Semi-annual cosine term & $\cos_{6,t} = \cos(\frac{2\pi \cdot t}{6})$\\ 
  year & number of years since start of timeseries (2015)&\\
  $\text{License}$& Number of active SRFS fishing licenses in a given month & Florida Fish and Wildlife Conservation Commission \\
  Unfishable & Fraction of unfishable days in a month & Global Surface Summary of the Day (`GSOD') weather data from the from the USA National Centers for Environmental Information \citep[`NCEI';][]{GSODR2018}\\
  $\text{Fuel}$ & Florida mean gasoline price (adjusted for inflation using 2024 dollars) & \href{https://www.eia.gov/dnav/pet/hist_xls/EMM_EPMRU_PTE_SFL_DPGm.xls}{United States Energy Energy Information Administration}\\ 
\botrule
\end{tabular*}
\end{table*}

%\FloatBarrier

For highly targeted species, anglers often compensate for shorter fishing seasons by increasing the number of trips taken during the open season, a phenomenon known as effort compression \citep{powers2016estimating, powers2018compression, topping2019comparison, trudeau2022lower}. As a result, recreational harvest does not necessarily decline linearly with shorter seasons. To account for this nonlinearity, we included the duration of the gag season (for both federal and statewide waters) as a predictor of harvest rate. Building on the logic presented in \cite{Hyman2024Effort}, we used the natural logarithm of season duration as a predictor. This approach reflects the principle that changes in season duration have a greater relative impact on harvest rates when the season is short. In contrast, as the season lengthens, the marginal impact of additional days diminishes, as longer seasons often exceed the practical number of fishing opportunities most anglers can realistically pursue.

To arrive at pre-season and in-season projections, NOAA's Southeast Regional Office (SERO) commonly uses an averaging approach, where the projected harvest for a specific month is calculated using the average harvest rate -- both SRFS and FHS combined divided by the number of days open to harvest -- observed during the same month over the three most recent years when that month was open to harvest. In instances when the recreational season is closed, the harvest rate is assumed to be zero. Although this method attempts to account for seasonal trends and variability, the accuracy of this approach is predicated on an assertion that past conditions underlying fishing conditions have remained unchanged -- a strong assumption which does not always hold \citep{carter2015nowcasting}. To compare this approach to more complex modeling frameworks, we included this historical average (denoted ``$\text{Past}$'') as a predictor in all models. The simplest tested model predicted harvest based only on this predictor and our gag temporal regulation predictors. However, for years 2015, 2016, 2017, SRFS data were not available to calculate this three-year average since the survey began in 2015. To obtain three-year averages for these years, we supplemented SRFS data with private-recreational data from APAIS weighted by the MRIP \href{https://www.fisheries.noaa.gov/recreational-fishing-data/fishing-effort-survey-glance}{Fishing Effort Survey (FES)} and calibrated to be consistent with SRFS units using established methodology \citep{Calibrating2021, Ramsay2024}. Finally, the four-county special season introduced complexities into the estimation process, as harvest rates during a month open only to these four counties differed from rates when all of Florida's Gulf Coast waters are open. To address this, we adjusted the three-year averaging method. Specifically, if a given month-year $t$ was open solely to the four-county season, we based the average on harvest rate data from the same month in the three most recent years when only the four-county season was active. Conversely, if the month-year $t$ was open to all state and federal waters, the average was derived from the three most recent years in which the entire Gulf Coast was open to recreational harvest of gag. This adjustment ensured that projections accounted for the distinct harvesting patterns associated with spatial variations in the gag season.

Red snapper is arguably the most highly targeted reef fish species by anglers in the Gulf, and consequently, red snapper temporal regulations can affect other species within the wider reef-fish complex. Recent evidence suggests that the status of the recreational red snapper season -- open or closed -- significantly impacts both recreational fishing effort overall and gag harvest specifically \citep{Hyman2024Effort, Hyman2025Catch}. This effect is likely driven by red snapper's popularity and the shared tendency of red snapper and gag to aggregate around structured reef habitats. To account for this dynamic, we included the proportion of time red snapper harvest was permitted in a given month as a predictor in our models. 

Seasonality influences both angler behavior and gag vulnerability to harvest. Recreational angler effort fluctuates throughout the year, often driven by environmental conditions, social events, and other external factors \citep{farmer2015forecasting,farmer2020forecasting,trudeau2022lower,Hyman2024Effort}. Additionally, seasonal movements associated with gag life history, such as spawning migrations and habitat use, can alter their susceptibility to fishing pressure \citep{gruss2017ontogenetic,heyman2019cooperative,lowerre2020testing}. To capture these seasonal dynamics, regression models can incorporate harmonic terms, such as sine and cosine functions, which account for periodic variations on annual or semi-annual cycles \citep[e.g.,][]{trudeau2022lower,Hyman2024Effort}. In our models, we included four harmonic terms: sine and cosine functions oscillating at 12-month (annual) and 6-month (semi-annual) frequencies, enabling a more precise representation of seasonal effects.

We considered an annual trend as a predictor of gag harvest rates. The inclusion of an annual trend in recreational fishing harvest models accounts for long-term changes such as improved catch efficiency from advancements in technology and gear \citep[e.g.,][]{marchal2007impact, selgrath2018shifting, detmer2020fishing}, as well as shifts in angler behavior and environmental conditions. These trends help distinguish overarching patterns from short-term fluctuations, enhancing model accuracy.

Recreational fishing activity is often influenced by weather conditions, with anglers less likely to fish during inclement weather \citep{farmer2020forecasting}. To evaluate this relationship, we used the \texttt{GSODR} package in R to generate daily weather summaries along the Gulf Coast of Florida for each day and month between 2015 and 2024 \citep{GSODR2018}. Weather data were extracted from all available USA National Centers for Environmental Information stations located within 30 km of the Gulf Coast and within the study period. A total of 37 stations met these criteria. We defined days with average wind speeds exceeding 7.5 m/s as ``unfishable'' \citep{farmer2020forecasting}. For each month, we calculated the proportion of unfishable days, denoted as the variable ``unfishable''. This metric was included as a covariate in our analyses to account for the impact of adverse weather on recreational fishing effort and harvest.

Finally, we considered two socioeconomic variables -- the number of Florida state reef fish license holders and annual fuel prices -- as predictors of gag harvest. Beginning in 2015, Florida required the Gulf Reef Fish Angler (GRFA) designation -- later adapted to the State Reef Fish Angler (SRFA) designation -- for licensed recreational fishermen to target select fish within the broader multispecies reef-fish complex. We hypothesized that the number of anglers with the GRFA or SRFA designation in each month may be related to fishing effort targeted towards reef fish, and by extension, gag harvest. Moreover, as fuel prices may affect anglers' decisions to fish and/or the length of fishing trips, we considered monthly fuel prices -- adjusted for inflation -- as an economic predictor.

\subsection{Analysis} \label{Analysis}
Monthly recreational gag harvest rates (landings per day open) were analyzed using a hurdle-gamma (HG) generalized linear model. This model has a probability mass function specified separately for zero and a probability density function for nonzero (positive) outcomes. The conditional density function for the HG probability distribution is used to describe the distributional forms:

\begin{flalign}
\text{HG}(y|\alpha, \phi, \theta) = \begin{cases} 
\theta, &y = 0\\
(1-\theta)\text{Gamma}(y|\alpha, \phi),&y \ge 0
\end{cases}
\end{flalign}

\noindent where $\theta$ is the probability of observing a zero and $\text{Gamma}(y|\alpha, \phi)$ is the Gamma probability density function for the gag harvest rate with inverse-scale and shape parameters $\alpha$ and  $\phi$. Notably, these parameters can be re-expressed in terms of the mean ($\mu$) and variance ($\sigma^2$). We therefore employed a distributional regression approach where we modeled the mean and variance directly. Consequently, the harvest rate for gag is expressed as:

\begin{flalign}
y_{t} &\sim \text{HG}(\alpha_{t}, \phi_{t}, \theta_{t})\\
& \alpha_{t} = \frac{\mu_{t}}{\sigma^2_{t}}\nonumber\\
& \phi_{t} = \frac{\mu_t^2}{\sigma_t^2}\nonumber\\
& \mu_{t} = exp\bigg(\beta_0 + \sum_{i=1}^Ix_{t,i}\beta_{i}\bigg)\nonumber\\
& \sigma^2_{t} =  exp\bigg(\rho_0 + \sum_{j=1}^J{z_{t,j}}\rho_{j}\bigg)\nonumber\\
& \theta_{t} = \frac{1}{1+exp(\lambda_0 + \sum_{l=1}^L{k_{t,l}\lambda_{l}})}\nonumber\\
&\qquad \beta, \rho, \lambda \sim N(0, 4)\nonumber
\end{flalign}

\noindent where $y_{t}$ denotes the weight of gag harvested per day open in the $t^{th}$ month-year. Meanwhile, $\mu_{t}$ and $\sigma^2_t$ denote the mean and variance of the Gamma distribution conditioned on $y_{t} > 0$, respectively, and $\theta_{t}$ denotes the binomial probability of observing 0 gag caught in month-year $t$. We use $x$, $z$, and $k$ to refer to predictors included in modeling $\mu_t$, $\sigma^2_t$, and $\theta_t$, respectively. Similarly, coefficients corresponding to $\mu_t$, $\sigma^2_t$, and $\theta_t$ are denoted $\beta$, $\rho$, and $\lambda$, respectively. Both $\mu_{t}$ and $\sigma^2_t$ are related to their predictors and associated coefficients through a log-link function, whereas $\theta_t$ is related to its predictors and associated coefficients through a logit-link function. All regression coefficients were assigned relatively uninformative, normally-distributed priors with mean 0 and variance 16.

We used the Stan probabilistic programming language for Bayesian modeling \citep{gelman2015stan}, accessed via R \citep{Rcore2024}, to fit all models. Each model involved four parallel Markov chains with 5,000 iterations for warm-up and another 5,000 for posterior sampling, yielding 20,000 total draws for inference. Covariates were deemed significant if their 80\% confidence intervals excluded 0 \citep{ipccAR6, kruschke2021bayesian, van2021bayesian}. All graphics and auxiliary analyses were produced using the \texttt{ggplot2} package in R \citep{ggplot2}.

\subsubsection{Model selection} \label{Selection}
We constructed five candidate models to predict Gulf gag harvest rates. For all models, we employed the same predictors to estimate the hurdle and variance components and examined the predictive performance of differing mean structures (Table \ref{tab2}). For our simplest model ($g_1$), we considered gag temporal regulations and the average historical gag harvest as predictors of mean gag harvest for the current year. This model was designed to emulate the methodology currently employed to predict harvest while also taking into account historic subtleties associated with gag temporal regulations across the Gulf Coast of Florida. Our second model, $g_2$, expanded on model $g_1$ to include an annual trend, seasonal harmonic terms, and red snapper temporal regulations. Model $g_3$ expanded on model $g_2$ to include social and economic variables, including the number of state reef-fish license holders in each month as well as the average cost of fuel, adjusted for inflation. Conversely, model $g_4$ expanded on model $g_2$ to include a weather variable, namely the fraction of each month not considered fishable. Finally, model $g_5$ served as our ``global'' model, which included all predictors considered. For more information on predictors used to estimate the hurdle and variance components, see Table \ref{tab3}.

\begin{table}[ht]
\caption{Fixed effects formulas for the mean structure of the five candidate models considered. All models had identical fixed effect formulas for modeling the shape and hurdle parameters. \label{tab2}}%
\begin{tabular*}{\columnwidth}{@{\extracolsep\fill}p{2cm} p{6cm}@{\extracolsep\fill}}
\toprule%
\multicolumn{1}{l}{\textbf{Model}} &\multicolumn{1}{l}{\textbf{Formula}}  \\
\hline
$g_1$ & $\text{Closed}_{Gag}$ + $\text{Special}_{Gag}$ + $\text{Full}_{Gag}$ + $\ln \text{Past}$ \\
  $g_2$ & $\text{Closed}_{Gag}$ +  $\text{Special}_{Gag}$ + $\text{Full}_{Gag}$ + $\ln \text{Past}$ + $\sin_{12}$ + $\cos_{12}$ + $\sin_{6}$ + $\cos_{6}$ + $\text{year}$ + $\text{Days}_{RS}$ \\ 
  $g_3$ &$\text{Closed}_{Gag}$ +  $\text{Special}_{Gag}$ + $\text{Full}_{Gag}$ + $\ln \text{Past}$ + $\sin_{12}$ + $\cos_{12}$ + $\sin_{6}$ + $\cos_{6}$ + $\text{year}$ + $\text{Days}_{RS}$ + $\text{License}$ + $\text{Fuel}$\\ 
  $g_4$ & $\text{Closed}_{Gag}$ +  $\text{Special}_{Gag}$ + $\text{Full}_{Gag}$ + $\ln \text{Past}$ + $\sin_{12}$ + $\cos_{12}$ + $\sin_{6}$ + $\cos_{6}$ + $\text{year}$ + $\text{Days}_{RS}$ + $\text{unfishable}$\\ 
  $g_5$ & $\text{Closed}_{Gag}$ +  $\text{Special}_{Gag}$ + $\text{Full}_{Gag}$ + $\ln \text{Past}$ + $\sin_{12}$ + $\cos_{12}$ + $\sin_{6}$ + $\cos_{6}$ + $\text{year}$ + $\text{Days}_{RS}$ + $\text{License}$ + $\text{Fuel}$ + $\text{unfishable}$\\
\botrule
\end{tabular*}
\end{table}

\begin{table}[ht]
\caption{Fixed effects formulas for the hurdle and variance structure of all models considered. \label{tab3}}%
\begin{tabular*}{\columnwidth}{@{\extracolsep\fill}ll@{\extracolsep\fill}}
\toprule%
\multicolumn{1}{l}{\textbf{Parameter}} &\multicolumn{1}{l}{\textbf{Formula}}  \\
\hline
$\theta$ & $\text{Open}_{Gag}$ \\ 
$\sigma^2$ & $\sin_{12}$ + $\cos_{12}$ + $\sin_{6}$ + $\cos_{6}$ \\
\botrule
\end{tabular*}
\end{table}

We employed estimated log-pointwise predictive density (ELPD) and corresponding $\Delta_{ELPD}$ values to assess the predictive performance of each model within our set of statistical models $g_{i}$ \citep{vehtari2017practical}. ELPD values are commonly used to evaluate out-of-sample predictive accuracy, while $\Delta_{ELPD}$ values represent the difference between the ELPD of a given model and that of the best-performing model in the set. The ELPD and $\Delta_{ELPD}$ values were estimated using the approximate Leave-One-Out Information Criterion \citep[LOO-IC; ][]{gelman2015stan, vehtari2017practical}, with the calculations performed through the \texttt{loo} package \citep{loo}. When two models showed similar $\Delta_{ELPD}$ values \citep[i.e., $\le$4;][]{sivula2020uncertainty}, the simpler model was selected based on the principle of parsimony.

\subsubsection{Cross-validation}\label{CV}
We employed extensive cross-validation (CV) to assess the predictive power of our best-performing statistical model. We considered two CV exercises -- block-CV and 1-step-ahead-performance-CV (1-SAP-CV) -- to separately assess the suitability of our model for pre-season and in-season projections, respectively. For our block-CV exercise, we refit the model to truncated datasets eight times -- each with one year withheld -- and subsequently compared predictions derived from our models to withheld, observed response variables for all months in the withheld year. As managers would only have information from previous years when making pre-season projections, this cross-validation method was a reasonable approach to assessing model pre-season posterior prediction accuracy. In contrast, nowcasting entails making in-season projections for the current time-step based on previous information both from the current year and previous years \citep{carter2015nowcasting}. As a result, for 1-SAP CV, for each year, we subsequently refit the model 11 times (once for each month). For each refit, we included months 1 to $m$ in our training set and then compared posterior predictive distributions to observed data in month $m+1$.  

For each posterior scan ($s$) of 1,000 used, monthly harvest rates were converted to total harvest by multiplying the predicted rates by the number of days the month was open to harvest. In addition, for block-CV we then calculated the cumulative harvest for month $m$ as the sum of harvest estimates across months ($1, \dots, m$) using posterior predictive scans derived from the hurdle-gamma distribution. To evaluate the practical utility of our approach, posterior predictions of both individual monthly harvest (using 1-SAP-CV) and cumulative harvest (using block-CV) were then compared to withheld data. This comparison assessed out-of-sample performance and demonstrated the applicability of our projections in real-world management scenarios.

\subsection{Comparison to current methods}
We evaluated the predictive performance of our model, both in-sample and out-of-sample, against estimates derived using historical average harvest. To mirror current prediction methods, we multiplied the historical average harvest rate by the number of days open to gag harvest in the current month to estimate monthly harvest values. These estimates were compared to those from our model under three scenarios: (1) fitted to the full dataset, (2) fitted to datasets with individual years withheld (block-CV), and (3) fitted to datasets with individual months withheld (1-SAP-CV), alongside observed data. For block-CV, we applied our pre-season cumulative estimation procedures to the historical average harvest rates to enable direct comparisons with our modeling approach. Comparative predictive performance was assessed based on the root-mean-square error (RMSE) of each estimate (from our model [i.e., the median posterior predictive] or historical average harvest) relative to observed harvest values.

\subsubsection{2025 Projection}\label{Projection}
After establishing the posterior predictive accuracy of our models, we conducted simulations to predict the 2025 gag season duration using 1,000 posterior scans from our best-fitting model. To begin, we calculated the adjusted annual catch target (ACT) for 2025 by accounting for the 2024 overage penalty—defined as the amount of gag harvested in 2024 (249,000 lbs) that exceeded the 2024 ACL (163,000 lbs). This overage was subtracted from the 2025 ACT (319,000 lbs) specified in the rebuilding plan \citep{Amendment56, NOAA2024GagSeason, ECFR2025}:  

\begin{flalign}
\text{2024 Penalty} &= \text{2024 ACL} -  \text{2024 Harvest}\nonumber\\
&= 163,000~\text{lbs}-  249,000~\text{lbs} = -86,000~\text{lbs}\nonumber\\
\text{2025 ACT} &= \text{2025 unadjusted ACT} + \text{2024 Penalty}\nonumber\\
&= 319,000~\text{lbs} - 86,000~\text{lbs} = 233,000~\text{lbs}\nonumber
\end{flalign}

To develop our simulations, we constructed a design matrix where all predictors, except for gag temporal regulations and seasonal variables, were fixed at their 2024 values. We began by setting the season duration to $d = 1$ day, with a starting date of September 1\textsuperscript{st}. Months without any days open to harvest were assigned the ``Closed'' temporal regulation treatment, whereas months with days open were assigned the ``Full'' treatment. Using our best-fitting model, we simulated posterior-predictive distributions of monthly harvest rates for 2025 across 1,000 posterior scans ($s$). For each scan, monthly harvest rates were multiplied by the number of days open to harvest within the corresponding month to derive posterior predictions of monthly harvest. To estimate total harvest for each scan ($s$), we summed harvest estimates across all months. The process was repeated iteratively, incrementing the season duration ($d$) by one day at a time, until $d=121$, which corresponded to the maximum possible season duration ending December 31\textsuperscript{st}. For each season duration ($d = 1, \dots, 121$), we calculated the median and 80\% posterior predictive interval of total harvest. Additionally, for each season duration, we estimated the probability that the ACT would be met or exceeded as the proportion of posterior predictive scans whose total harvest met or exceeded the ACT. Finally, we compared these projections to those using the historic average harvest rate for September.

\section{Results}
\subsection{Model selection and diagnostics}
The best-fitting model was $g_{2}$, which described mean gag harvest as a function of gag temporal regulations, past harvest, seasonal terms, an annual trend, and red snapper temporal regulations. While models $g_3$, $g_4$, and $g_5$ demonstrated similar predictive performance, with $\Delta_{ELPD} \leq 4$ (Table \ref{tab4}), model $g_2$ was the simplest among the four. Consequently, we selected model $g_2$ for inference, adhering to the principle of parsimony. Importantly, the ELPD values for models $g_2$, $g_3$, $g_4$, and $g_5$ were all substantially higher than that of model $g_1$, indicating that incorporating additional predictors beyond gag temporal regulations and past harvest significantly enhanced predictive power. Herein, all inferences are based on model $g_2$.

\begin{table}[ht]
\caption{Model selection results from five Bayesian hurdle-gamma regression models ($g_{i}$) predicting recreational gag harvest. Models are presented in order of predictive power based on collected data. \textbf{LOO}: the approximate Leave-One-Out Information Criterion; \textbf{ELPD$_{\text{LOO}}$}: the estimated log-pointwise density calculated from LOO; \textbf{$\bm{\Delta}_{\textbf{ELPD}}$}: the relative difference between the ELPD of any model and the best model in the set; \textbf{SE}$_{\bm{\Delta}_{\textbf{ELPD}}}$: standard error for the pairwise differences in ELPD between the best model and any given model. The selected model ($g_2$) values are presented in bold font. \label{tab4}}%
\begin{tabular*}{\columnwidth}{@{\extracolsep\fill}lrrrr@{\extracolsep\fill}}
\toprule%
\multicolumn{1}{l}{\textbf{Model}} &\multicolumn{1}{l}{\textbf{$\text{LOO}$}} &\multicolumn{1}{l}{\textbf{$\text{ELPD}_\textbf{LOO}$}}&\multicolumn{1}{l}{$\bm{\Delta_{\textbf{ELPD}}}$}&\multicolumn{1}{l}{\textbf{SE}$_{\bm{\Delta_{\textbf{ELPD}}}}$}\\
\hline
$\bm{g_2}$ & \textbf{1448.61} & \textbf{-724.31} & \textbf{0} & \textbf{0} \\ 
  $g_4$ & 1450.83 & -725.42 & -1.11 & 0.73 \\ 
  $g_3$ & 1454.08 & -727.04 & -2.73 & 1.22 \\ 
  $g_5$ & 1455.08 & -727.54 & -3.23 & 1.50 \\ 
  $g_1$ & 1490.43 & -745.21 & -20.91 & 6.92 \\ 
\botrule
\end{tabular*}
\end{table}

\begin{figure*}[htbp]
\center{\includegraphics[width=\textwidth]{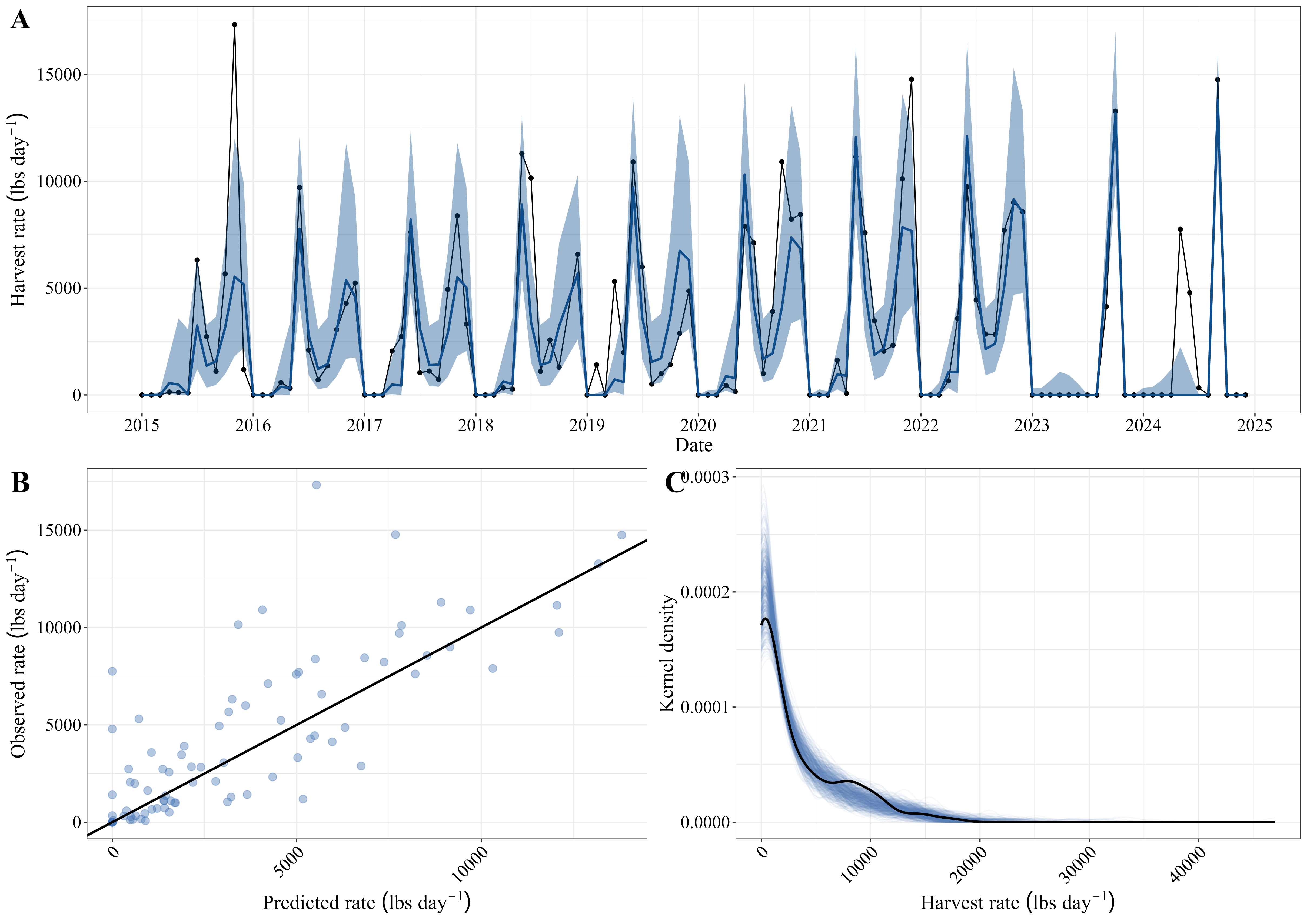}}\textbf{}
\caption{{Model fit and diagnostic plots for the best-fitting model ($g_2$).}
A: Observed (black) and expected (blue) monthly gag harvest rate. Bands denote upper and lower 80\% posterior predictive intervals. Points shaded black denote observations. B: Scatter-plot of median predicted versus observed gag harvest rate with the 1-1 line (black) superimposed. C: Comparison of the empirical distribution of observed gag harvest rate (black) to the distributions of 500 scans from the posterior predictive distribution (blue).}
\label{fig:Fig1}
\end{figure*}

Time-series plots comparing observed and predicted values, along with posterior predictive checks, indicated that model $g_2$ provided a good fit to the observed data. The posterior predictive median and 80\% CI of monthly estimates aligned closely with in-sample estimates of monthly gag harvest rates (Fig. \ref{fig:Fig1}a). Additionally, diagnostics derived from model $g_2$ demonstrated satisfactory model performance (Figs. \ref{fig:Fig1}b and \ref{fig:Fig1}c).

\subsection{Predictors of monthly gag harvest rates}
\begin{figure}[htbp]
\center{\includegraphics[width=\columnwidth]{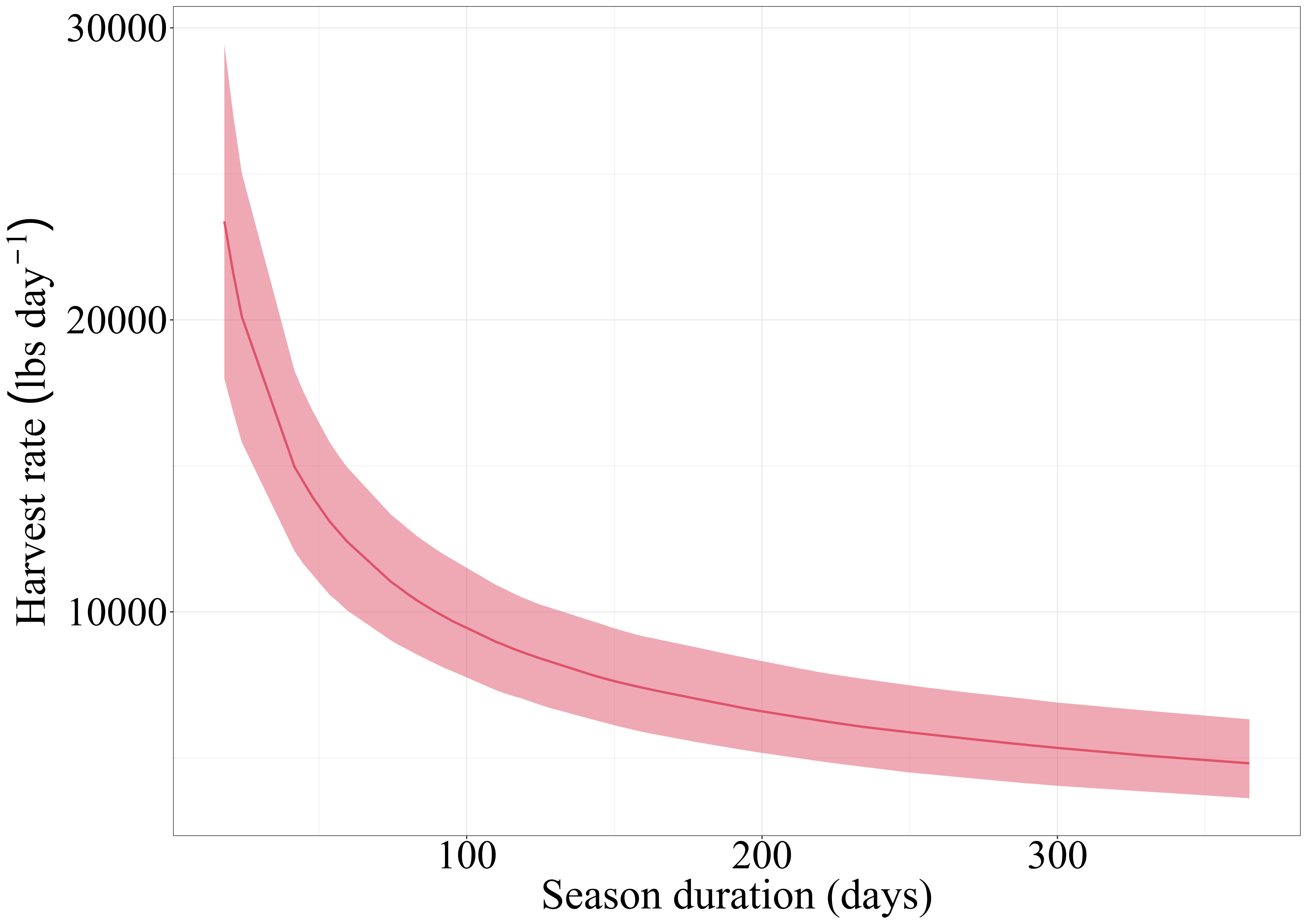}}\textbf{}
\caption{Conditional effects plots depicting the conditional expectation of harvest rate as a function of gag season durations. Colored lines and bands denote median and nominal 80\% Bayesian conditional prediction intervals derived from posterior predictive distributions with all other coefficients fixed at September 2024 values (i.e., the final month in the dataset open to harvest).}
\label{fig:Fig2}
\end{figure}

Gag temporal regulations, the annual trend, and seasonal harmonic terms all significantly explained variation in mean (nonzero) recreational gag harvest rates (Table \ref{tab5}). Conditioned on a positive outcome and relative to a closed season, months when gag was open to the entire Florida Gulf Coast (Full$_{Gag}$) resulted in higher harvest rates, while months open only for the four-county season (Special$_{Gag}$) resulted in lower harvest rates. The duration of the gag season had a clear negative effect on harvest rates, such that the harvest rate was lower during longer gag seasons (Fig. \ref{fig:Fig2}). The annual trend was positive and significant, indicating that after accounting for other predictors, gag harvest rates appeared to be increasing over time. Notably, after accounting for the effects of other predictors, past harvest did not significantly explain residual variation in gag harvest rates.

\begin{table}[h]
\caption{Posterior summary statistics (median and 80\% CIs) for coefficient estimates for predictors in mean ($\mu$), hurdle ($\theta$), and variance ($\sigma^2$) components in the best-fitting gag harvest model ($g_2$). Coefficient terms with asterisks denote 80\% CIs that excluded 0.  \label{tab5}}%
\tabcolsep=0pt%%
\begin{tabular*}{\columnwidth}{@{\extracolsep{\fill}}llrrr@{\extracolsep{\fill}}}
\toprule%
\multicolumn{1}{l}{\textbf{Component}} &\multicolumn{1}{l}{\textbf{Predictor}} &\multicolumn{1}{l}{\textbf{10}\%}&\multicolumn{1}{l}{\textbf{50}\%}&\multicolumn{1}{l}{\textbf{90}\%}\\
\multicolumn{1}{l}{} &\multicolumn{1}{l}{\textbf{(Coefficient)}} &\multicolumn{1}{l}{}&\multicolumn{1}{l}{}&\multicolumn{1}{l}{}\\
\hline

$\mu$ & $\text{Closed}_{Gag}$ ($\beta_{0}$) & 8.28 & 9.64 & 10.92\\ 
   & $\text{Special}_{Gag}$* ($\beta_{1}$) & -1.85 & -1.31 & -0.62\\ 
   & $\text{Full}_{Gag}$* ($\beta_{2}$) & 0.50 & 0.84 & 1.24\\ 
   & $\text{Season}_{Gag}$* ($\beta_{3}$) & -0.64 & -0.53 & -0.39\\ 
   & $\ln{\text{Past}}$ ($\beta_{4}$) & -0.09 & 0.04 & 0.18\\ 
   & $\sin_{12}$* ($\beta_{5}$) & 0.10 & 0.37 & 0.60 \\ 
   & $\sin_{6}$* ($\beta_{6}$) & -0.85 & -0.64 & -0.47 \\ 
   & $\cos_{12}$ ($\beta_{7}$)& -0.14 & 0.08 & 0.28\\ 
   & $\cos_{6}$* ($\beta_{8}$) & -0.38 & -0.23 & -0.06\\ 
   & $\text{Year}$* ($\beta_{9}$) & 0.05 & 0.08 & 0.11\\ 
   & $\text{Days}_{RS}$ ($\beta_{10}$) & -0.13 & 0.18 & 0.48\\ 

  $\theta$ & ---* ($\lambda_{0}$) & 1.64 & 2.23 & 2.91\\ 
   & $\text{Open}_{Gag}$* ($\lambda_{1}$) & -10.50 & -7.89 & -6.16\\ 

  $\sigma^2$ & ---* ($\rho_{0}$) & 14.19 & 14.47 & 14.78\\ 
   & $\sin_{12}$* ($\rho_{1}$) & -1.48 & -1.11 & -0.74\\ 
   & $\sin_{6}$* ($\rho_{2}$) & -2.10 & -1.65 & -1.16\\ 
   & $\cos_{12}$* ($\rho_{3}$) & -1.05 & -0.60 & -0.10\\ 
   & $\cos_{6}$ ($\rho_{4}$) & -0.37 & -0.02 & 0.33\\ 
\botrule
\end{tabular*}
\end{table}

The hurdle parameter, $\theta$ was strongly and negatively impacted by whether gag was open to harvest (either in the four-county region or across the Gulf Coast of Florida). When the season was completely closed to gag harvest, the probability of observing 0 gag harvested in a given month was 0.90 (80\% CI: 0.84 -- 0.95). However, this probability decreased to nearly 0 when gag was open to harvest. Meanwhile, both sinusoidal harmonic terms significantly explained $\sigma^2$, suggesting that model uncertainty varied seasonally.

\subsection{Comparison to historical average harvest}
\begin{figure*}[ht]
\center{\includegraphics[width=\textwidth]{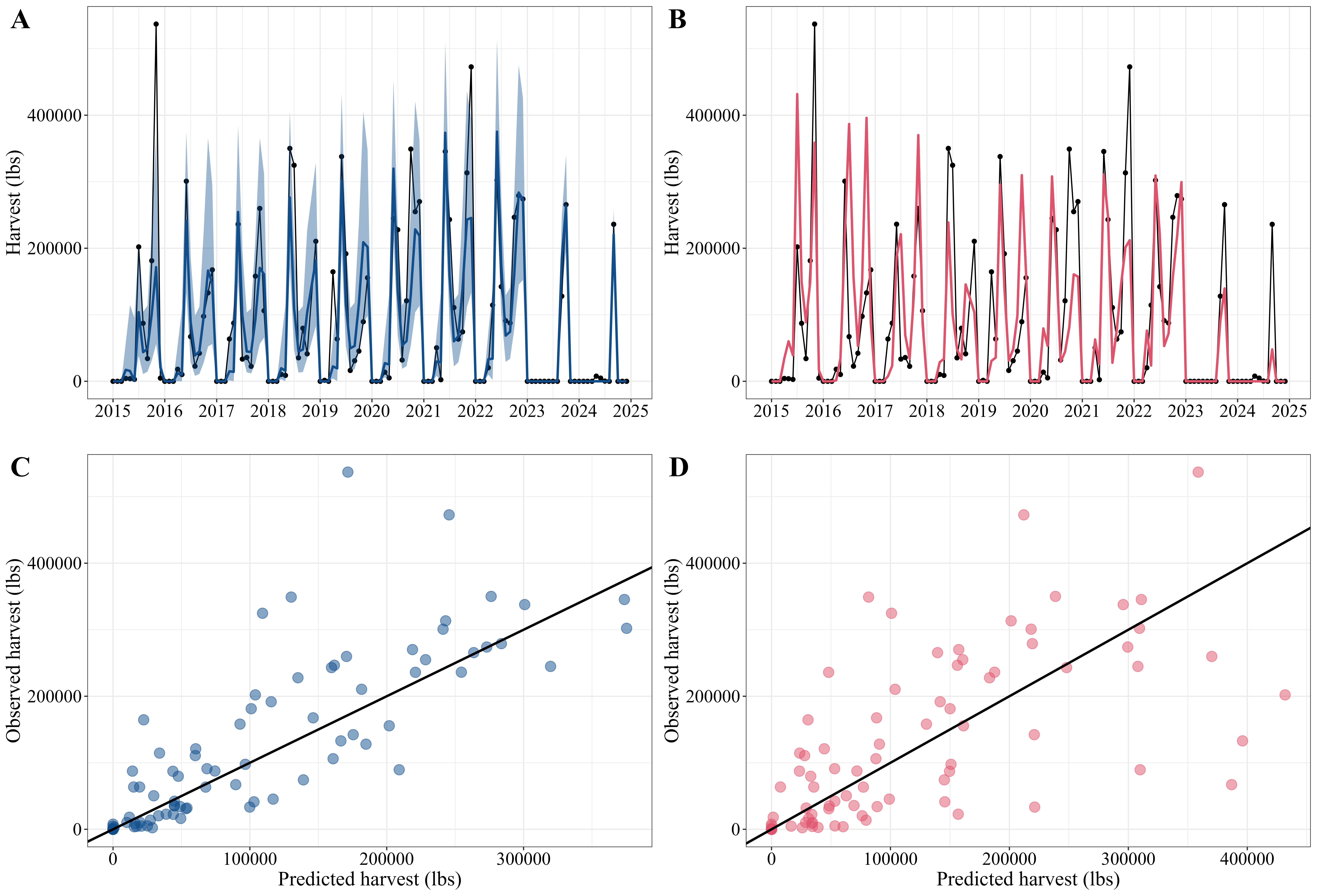}}
\caption{Comparison of in-sample predictive performance between monthly gag harvest estimates derived from model $g_2$ and historical average harvest. A: Observed (black) and expected (blue) monthly gag harvest using model $g_2$. Bands denote upper and lower 80\% posterior predictive intervals. Points shaded black denote observations. B: Observed (black) and expected (red) monthly gag harvest using historical average harvest. C: Scatter-plot of median predicted versus observed gag harvest with the 1-1 line (black) superimposed using model $g_2$. D: Scatter-plot of median predicted versus observed gag harvest with the 1-1 line (black) superimposed using historical average harvest.}
\label{fig:Fig3}
\end{figure*}

Comparisons of monthly harvest estimates derived from the in-sample fit of model $g_2$ with monthly harvest estimates derived using the historical average harvest demonstrated the superior performance of model $g_2$ (Fig. \ref{fig:Fig3}). The $R^2$ value between the model $g_2$ posterior predictive median estimates and observed harvest was 0.75 (Fig. \ref{fig:Fig3}C), compared to 0.54 using historical average harvest (Fig. \ref{fig:Fig3}D). In addition, the RMSE between observed harvest and median estimates derived from model $g_2$ was 79,000 lbs, compared to 105,000 lbs (i.e., 25\% improvement). Notably, for years 2023 and 2024, model $g_2$'s RMSE was 57\% (40,000 vs. 93,000 lbs) and 92\% (15,000 vs. 188,000 lbs) lower than RMSE from historical average harvest.

\subsection{Cross-validation}
\begin{figure*}[htbp]
\center{\includegraphics[width=\textwidth]{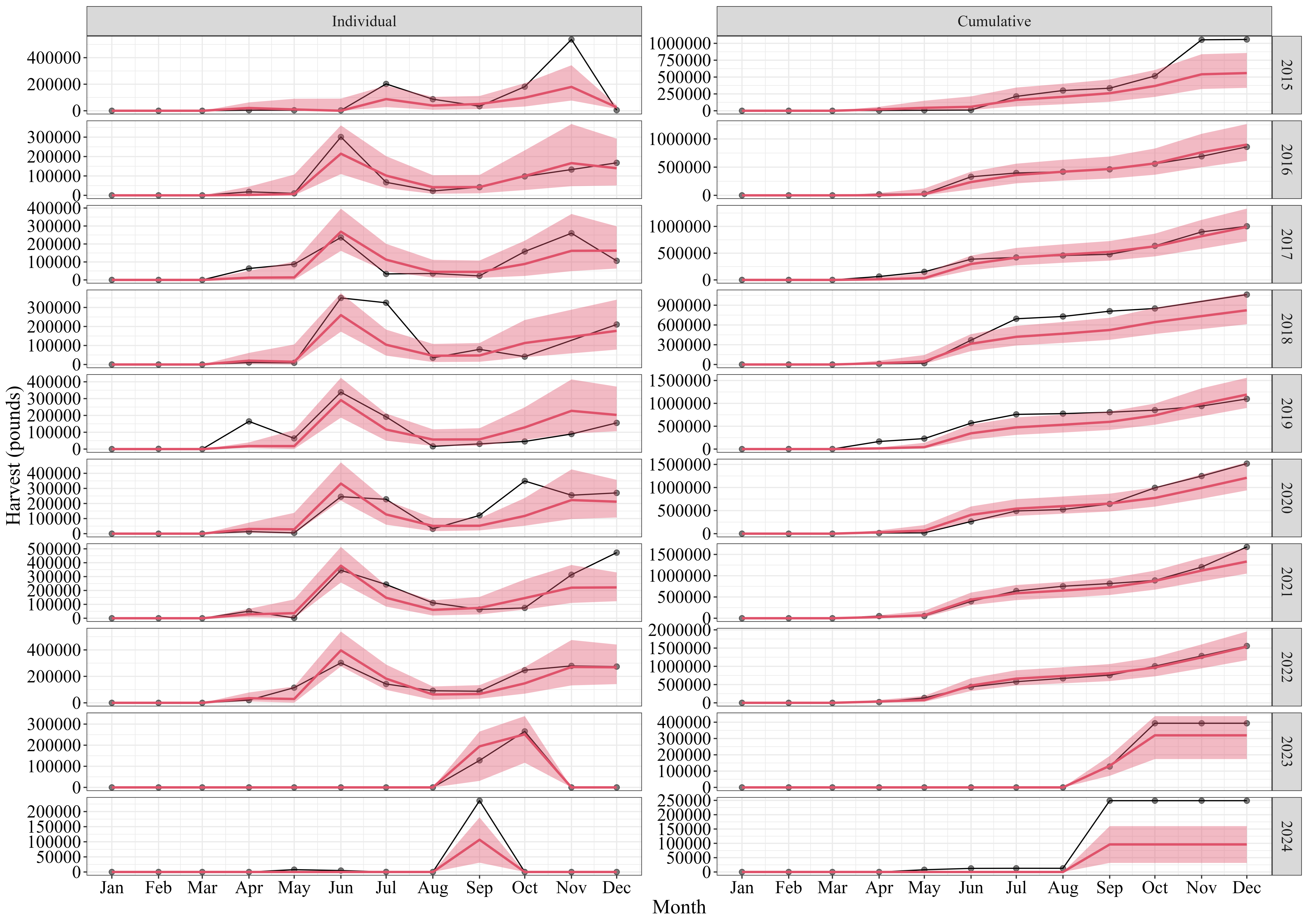}}\textbf{}
\caption{Individual monthly and cumulative cross-validation (CV) results for the best-fitting model ($g_2$).Model posterior predictive intervals for each year are derived from refitted versions of model $g_2$ on truncated datasets that year withheld (i.e., out-of sample). Black points denote observed harvest, the red line denotes the expected gag harvest, while bands denote upper and lower 80\% posterior predictive intervals. Left: Individual monthly values estimated using 1-SAP-CV. Right: Cumulative pre-season values estimated using block-CV. Note that each y-axis is plot-specific.}
\label{fig:Fig4}
\end{figure*}

Cross-validation suggested our best-fitting model was statistically robust and reliably predicted withheld values. In addition, the model's median estimates outperformed the historical average harvest for both individual months and cumulative pre-season projections. For 1-SAP-CV, 83\% of withheld values were captured within 80\% posterior predictive intervals among all values (Fig \ref{fig:Fig4}). The average RMSE between the median prediction  using 1-SAP-CV and withheld values for individual months was 54,000 lbs, compared to an average RMSE of 66,000 lbs when using predictions derived from historical average harvest (i.e., an 18\% decrease in RMSE when switching from historical harvest to model $g_2$; Table \ref{tab:Tab6}).

\begin{table}[ht]
\centering
\caption{
Comparisons of average RMSE (lbs) between individual and cumulative monthly gag harvest values across all years from our best-fitting model using block-CV and 1-SAP-CV and historical harvest (``Historic''). Values are rounded to the nearest thousand pounds.}
\vspace{0.2cm}
\begin{tabular}{crrrr}
  \hline
  \multicolumn{1}{l}{\textbf{}} &\multicolumn{2}{c}{\textbf{Individual}} &\multicolumn{2}{c}{\textbf{Cumulative}}\\
\multicolumn{1}{l}{\textbf{Month}}&\multicolumn{1}{l}{\textbf{1-SAP}} &\multicolumn{1}{l}{\textbf{Historic}}&\multicolumn{1}{l}{\textbf{Block}}&\multicolumn{1}{l}{\textbf{Historic}}\\
\hline
  Jan & 0 & 0 & 0 & 0 \\ 
  Feb & 0 & 0 & 0 & 0 \\ 
  Mar & 0 & 0 & 0 & 0 \\ 
  Apr & 51,000 & 55,000 & 52,000 & 55,000 \\ 
  May & 41,000 & 45,000 & 76,000 & 81,000 \\ 
  Jun & 60,000 & 58,000 & 98,000 & 116,000 \\ 
  Jul & 98,000 & 150,000 & 131,000 & 185,000 \\ 
  Aug & 28,600 & 55,000 & 125,000 & 213,000 \\ 
  Sep & 54,000 & 71,000 & 130,000 & 236,000 \\ 
  Oct & 96,000 & 110,000 & 125,000 & 240,000 \\ 
  Nov & 136,000 & 142,000 & 205,000 & 287,000 \\ 
  Dec & 86,000 & 102,000 & 236,000 & 315,000 \\ 
  &&&&\\
  Average & 54,000 & 66,000 & 98,000 & 144,000 \\  \botrule
\end{tabular}\label{tab:Tab6}
\end{table}

For block-CV, across all withheld blocks (years), 78\% of cumulative values were captured within the 80\% posterior predictive interval (Fig. \ref{fig:Fig4}). For block-CV, the average RMSE between the model's median prediction for cumulative harvest and withheld values was 98,000 lbs (Table \ref{tab:Tab6}). In contrast, average RMSE between cumulative estimates using historical average harvest and observed harvest was 144,000 lbs (i.e., a 32\% average decrease in RMSE when shifting from current methods to model $g_2$). Notably, for each month, average RMSE of cumulative estimates was consistently equal to or lower than the corresponding RMSE from average values estimated using historical average harvest.

\subsection{Projections}
\begin{figure*}[ht]
\center{\includegraphics[width=\textwidth]{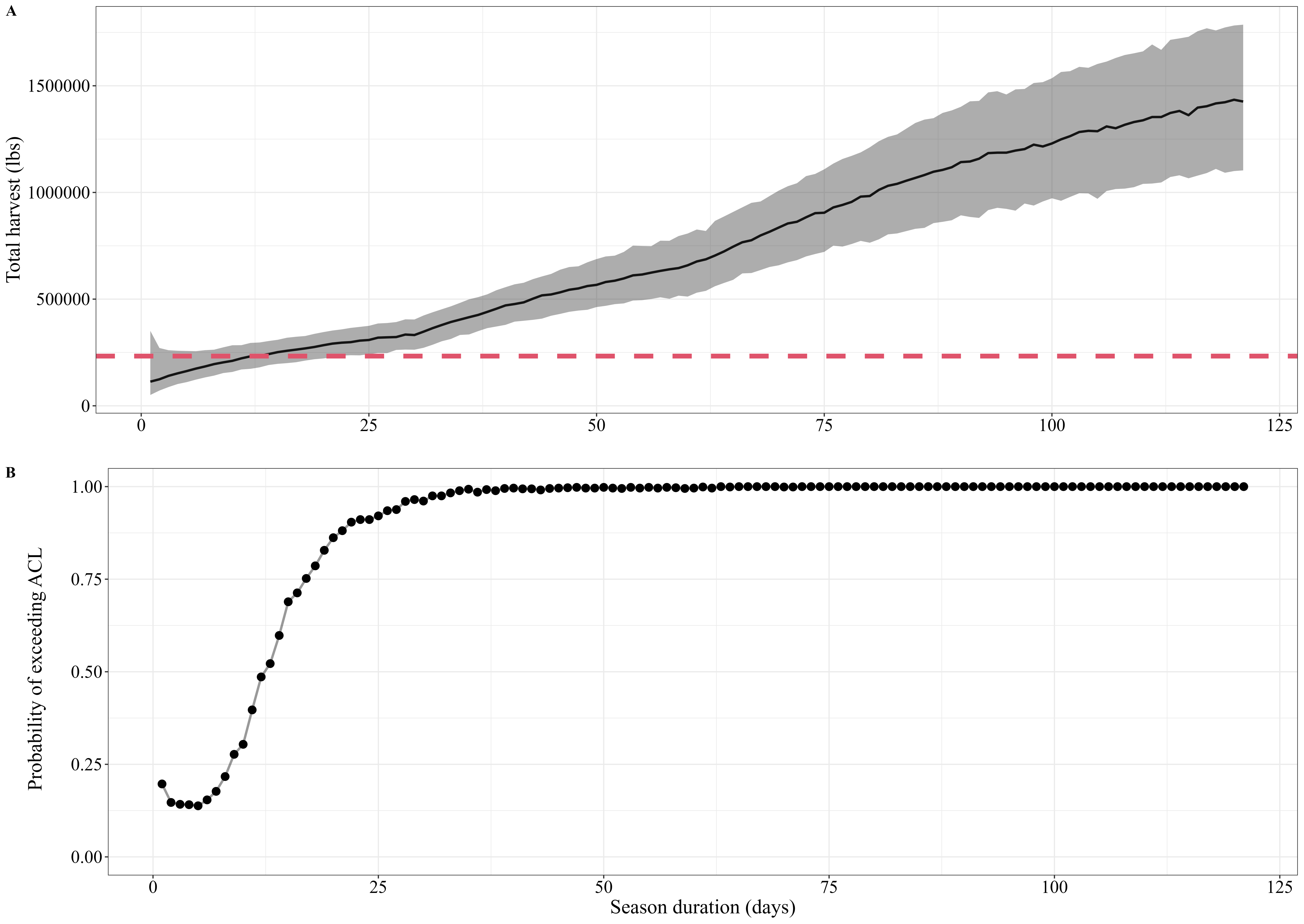}}\textbf{}
\caption{ Season duration projections for a 2025 gag season given a hypothetical annual catch target (ACT). (A) Counterfactual plot depicting the total gag harvested as a function of season duration for seasons ranging from one day to 121 days. The black line denotes the median total harvest estimate for each season duration, while the gray band denotes the 80\% CI. The red dashed line denotes a hypothetical 2025 ACT of 233,000 lbs. (B) Plot depicting the probability the hypothetical ACT would be exceeded for each possible season duration.}
\label{fig:Fig5}
\end{figure*}

Our simulations indicated that the hypothetical 2025 gag ACT of 233,000 lbs would likely be exceeded within approximately 12 days, with an 80\% confidence interval ranging from less than one day to 21 days (Fig. \ref{fig:Fig5}A). In contrast, a projection using average historical harvests would have resulted in a 29-day season. Simulations indicated a 20\% probability of exceeding the ACT even with a one-day season, reflecting the substantial uncertainty in harvest rate estimates during extremely short seasons (Fig. \ref{fig:Fig5}B). Meanwhile, the probability that harvest would exceed the ACT approached 100\% after 39 days. Simulations indicated the ACT would be exceeded by 1,800 lbs on day 13 (one day after the median projection), 58,000 lbs on day 20 (the upper 80\% confidence level), and 223,000 lbs by day 39.

\section{Discussion}
Effective management of recreational fisheries relies on accurately forecasting future harvests. In this study, we developed models to improve the prediction of gag harvests for both pre- and in-season management. Our results demonstrated that the best-fitting model enhanced forecasting accuracy, reducing cumulative pre-season projection errors by an average of over 32\% and in-season errors by almost 20\% compared to currently used methods across all years. However, more recent comparisons of observed harvest with estimates from in-sample model fits and historical harvest data revealed that historical harvest-derived estimates substantially underestimated observed values in 2023 and 2024. RMSE for model-derived estimates was 57\% and 92\% lower than for historical harvest-derived estimates in 2023 and 2024, respectively, suggesting that our statistical models are particularly well suited for predicting harvest following rapid restrictions on recreational seasons. Therefore, the ability of this framework to account for angler behavioral responses to reductions in season duration represents a notable improvement in management outcomes. These findings complement previous work highlighting the potential of statistical modeling approaches in setting initial season durations and monitoring harvest as the season progresses \citep{hanson2006forecasting, carter2015nowcasting, farmer2015forecasting, farmer2020forecasting, compaire2024time}. A second improvement of the proposed framework is its ability to explicitly quantify the probability of exceeding the harvest quota for any given season duration. As more data become available over time, the predictive performance of this framework—and similar approaches—is expected to improve, further increasing its value for fisheries management.

\subsection{Predictors of gag harvest}
Our results yielded insights into potential drivers of gag harvest. Gag temporal regulations, seasonal terms, and an annual trend all significantly explained variation in gag harvest rates, whereas past harvest, red snapper management regulations, the number of GRFA/SRFA license holders, fuel, and the fraction of unfishable days in a month did not. That gag temporal regulations significantly influenced gag harvest is unsurprising, as the choice to target a given fish species is dependent on whether that species is legally permitted to be harvested. Moreover, gag harvest rates were generally much higher when the season was open to all Florida counties along the Gulf than when the season was only open for the four-county miniature season in Franklin, Wakulla, Jefferson, and Taylor counties. This was expected, as the underlying population of these four counties -- and by extension the population of licensed anglers -- is a small fraction of all Florida counties along the Gulf. Furthermore, the significant influences of two out of the four harmonic terms in our model were consistent with expectations and previous work, as both recreational angler and gag behavior vary seasonally \citep{farmer2015forecasting, lowerre2020testing, Hyman2024Effort, Hyman2025Catch, compaire2024time}. Finally, the significant and positive annual trend may be related to increases in gear efficiency and changes in fishing effort directed at this species \citep[e.g.,][]{marchal2007impact, selgrath2018shifting, detmer2020fishing}. 

A key finding of this study is the pronounced effort compression observed in the recreational Gulf gag fishery. While previous research did not identify a relationship between gag season duration and recreational reef fishing effort \citep[e.g.,][]{Hyman2024Effort}, our model revealed a strongly negative effect on gag harvest rates, which increased non-linearly as season duration decreased. As noted by \cite{Hyman2024Effort}, a lack of contrast in season duration previously may have limited the detection of this effect. The recreational gag fishing seasons have only recently been significantly shortened, with the 2023 and 2024 seasons being the shortest on record. This is supported by our CV results: when 2024 data were included in model training, the model accurately predicted effort compression and correctly estimated withheld gag harvest rates for 2023 (Fig. \ref{fig:Fig4}). However, when 2024 data were excluded from the training dataset, the model underestimated harvest rates for 2024, suggesting that the compression effect was underrepresented. Although similar effort compression in response to shortened seasons has been documented for Gulf red snapper \citep[e.g.,][]{powers2016estimating, powers2018compression, topping2019comparison, farmer2020forecasting}, this study provides the first evidence of such effects in another important Gulf fishery. Given these findings, we recommend assessing harvest data for other vulnerable and highly targeted species with recently shortened seasons for evidence of effort compression, to better manage harvest quotas and minimize overfishing risks.

Somewhat surprisingly, several predictors hypothesized to influence gag harvest rates had no discernible effect. First, after accounting for other predictors, past monthly harvest averages did not significantly predict future gag harvest. Historical harvest data integrates influences such as angler effort, seasonal patterns, and gag abundance or availability. However, since many of these factors were explicitly addressed through other predictors in our model, the lack of significance suggests that the information captured by past harvest averages may already have been explained, rendering it redundant as a standalone variable. Similarly, we did not observe a positive relationship between gag harvest rates and red snapper temporal regulations, despite prior studies suggesting such an effect \citep{Hyman2025Catch}. A possible explanation for this discrepancy lies in the difference in spatial scale: whereas previous research divided Florida's Gulf Coast into multiple regions, our study aggregated data from the entire Florida Gulf Coast to align with the management approach for the Gulf gag stock. Since the impact of red snapper temporal regulations in earlier studies was region-specific, it is likely that aggregating the regions in our analysis obscured this effect.

\subsection{Predictive performance}
Despite utilizing relatively few predictors, our model represents a significant improvement in predictive performance compared to current methods. Cumulative projections of gag harvest, which are valuable for setting season durations before the season has started, were on average 32\% more accurate than current methods (mean difference in RMSE of 46,000 lbs). Given that the accuracy of future cumulative harvest estimates at the end of the season is crucial for accurately projecting season closure dates \citep[e.g.,][]{farmer2020forecasting}, our modeling framework offers significant advantages for determining gag season durations. Future gag seasons are expected to be more variable than those observed in the past decade due to the current rebuilding plan. As our model appears particularly more accurate than currently employed methods when seasons are highly restricted, it may be especially useful when stock size is low. Moreover, as season durations become more variable, the model’s predictive capacity will improve, benefiting from increased predictor contrast—an advantage not shared by historically used prediction methods. Monthly harvest predictions from the 1-SAP-CV model demonstrated significantly greater accuracy compared to the current methodology. This advancement holds critical implications for in-season management, where reporting delays, at present, often exceed 45 days \citep{carter2015nowcasting, MRIP2023}. Improved forecasting and nowcasting of harvest levels enable managers to refine projections of when seasonal quotas will be reached, thereby minimizing the risk of quota overages and the associated penalties to future allocations.

One of the most significant advantages of the proposed framework over current methods is its ability to explicitly quantify the risk associated with management action. As demonstrated in our 2025 simulation, this approach facilitates the estimation of both the probability of exceeding a specified harvest quota and the expected magnitude of any overage for each potential season duration. This risk-duration trade-off provides managers with a valuable tool to balance the benefits of longer seasons against the potential costs to stock health. Such precautionary, probabilistic approaches are particularly critical when managing vulnerable stocks with low biomass, where mismanagement can result in severe consequences, such as forced season closures in subsequent years or even stock collapse \citep{shertzer2010probabilistic, wilberg2019developing}.

\subsection{Future work}
While our modeling approach improves upon existing methodologies, future applications could enhance it in several key ways. First, our framework did not account for observation error inherent in the monthly estimates of gag harvest. While an earlier modeling attempt included these errors, we were unable to achieve model convergence. Incorporating observation error in future models could improve inference on influential predictors and enhance the accuracy of model predictions. Second, to simplify the modeling framework, we aggregated FHS for-hire and SRFS private-recreational gag harvest data into a single monthly harvest rate estimate. While this approach facilitated prediction, it may have constrained our ability to draw fleet-specific inferences. Future efforts should consider fleet-specific modeling strategies to identify fleet-specific drivers, particularly as more SRFS private-recreational data become available. Third, as mentioned in the Methods section, we excluded the shore mode from our analyses due to its relatively small contribution to gag landings and the high uncertainty associated with its estimates. However, NOAA SERO includes shore mode in its monitoring of the recreational gag quota, alongside the for-hire and private modes (now represented using SRFS). Therefore, future modeling efforts that incorporate all three modes may improve the accuracy of harvest projections. Finally, the relatively small sample size of our dataset may have limited the detection of effects from influential predictors. We recommend applying similar modeling frameworks to larger datasets or expanded predictor sets to improve accuracy and more robustly assess the relationships between harvest rates and explanatory variables.

\section{Competing interests}
No competing interests are declared.

\section{Author contributions statement}
All authors were influential in conceptualizing the study. A. C. Hyman, C. Ramsay, T. A. Cross and B. Sauls designed the methodology.  C. Ramsay and T. A. Cross performed data formatting/pre-processing. A. C. Hyman conducted all formal statistical analyses and was responsible for writing (original draft). T. K. Frazer provided resources to the project. All authors contributed to reviewing and critically revising the manuscript. All authors have approved the final version of the manuscript.

\section{Acknowledgments}
The authors thank M. Larkin and A. Gray from NOAA SERO for guidance on current methodology used to predict recreational gag seasons to ensure our depiction was accurate. This work was improved by  M. Drexler, whose helpful input led to constructive changes in the scope of our conclusions. This manuscript is a contribution from the Quantitative Fisheries Assessment Workgroup, a collaborative effort between the University of South Florida and the Florida Fish and Wildlife Research Institute.

\section{Funding}
Support for this work was provided, in part, by the Ocean Conservancy [grant number 2500189000], the Florida Fish and Wildlife Research Institute, and the University of South Florida.

\bibliographystyle{plain}
\bibliography{reference}
\end{document}